\documentclass[12pt]{iopart}
\usepackage{graphicx}
\usepackage{epstopdf}
\usepackage{amssymb}

\usepackage{color}
\usepackage[round]{natbib}
\newcommand{\newblock}{}
\usepackage{tabularx}
\usepackage{multirow}
\usepackage{booktabs}
\usepackage{array}
\usepackage{iopams}
\usepackage{changepage}

\usepackage{setstack}

\begin{document}
\bibliographystyle{natbib} 

\title[Spectrum estimation]{An indirect transmission measurement-based spectrum estimation method for computed tomography}

\author{Wei Zhao$^{1}$, Kai Niu $^{1}$, Sebastian Schafer$^{2}$, Kevin Royalty$^{2}$}%
\address{$^1$ Department of Medical Physics, University of Wisconsin-Madison, 1111 Highland Avenue, Madison, WI 53705}
\address{$^2$ Siemens Medical Solutions, USA Inc., Hoffman Estates, IL 60192}

\ead{wzhao49@wisc.edu}

\begin{abstract}

The characteristics of an x-ray spectrum can greatly influence imaging and related tasks. In practice, due to the pile-up effect of the detector, it's difficult to directly measure the spectrum of a CT scanner using an energy resolved detector. An alternative solution is to estimate the spectrum using transmission measurements with a step phantom or other CT phantom.  In this work, we present a new spectrum estimation method based on indirect transmission measurement and model spectra mixture approach. The estimated x-ray spectrum was expressed as weighted summation of a set of model spectra, which can significantly reduce the degrees of freedom (DOF) of the spectrum estimation problem. Next, an estimated projection can be calculated with the assumed spectrum. By iteratively updating the unknown weights, we minimized the difference between the estimated projection data and the raw projection data. The final spectrum was calculated with these calibrated weights and the model spectra. Both simulation and experimental data were used to evaluate the proposed method. In the simulation study, the estimated spectra were compared to the raw spectra which were used to generate the raw projection data. For the experimental study, the ground truth measurement of the raw x-ray spectrum was not available. Therefore, the estimated spectrum was compared against spectra generated using the SpekCalc software with tube configurations provided by the scanner manufacturer. The results show the proposed method has potential to accurately estimate x-ray spectra using the raw projection data. The difference between the mean energy of the raw spectra and the mean energy of the estimated spectra were smaller than 0.5 keV for both simulation and experimental data. Further tests show the method was robust with respect to the model spectra generator. 

\end{abstract}


\maketitle

\section{Introduction}
X-ray spectral characteristics greatly influence imaging and imaging related tasks such as Monte Carlo (MC) based radiation dose calculation, artefacts reduction techniques, and dual-energy material decomposition. In practice, it is difficult to directly measure the spectrum of a CT scanner with an energy resolved detector due to the limited count rate of the detector. Instead, indirect methods are widely used to estimate the spectrum of a CT scanner. These methods can be roughly classified into model-based methods which estimate the spectrum using empirical or semi-empirical physical models~\citep{birch1979,tucker1991,boone1997}, Compton-scattering measurement~\citep{yaffe1976,matscheko1987,matscheko1989compton,matscheko1989,gallardo2004,maeda2005}, transmission measurements~\citep{archer1982,hassler1998,waggener1999,sidky2005,zhang2007,xu2009,duan2011,lin2014} and MC simulation~\citep{llovet2003,ay2004,mainegra2006,bazalova2007,miceli2007a}. 


The aim of this work was to develop an indirect transmission measurement-based spectrum estimation method using only raw projection data of a physical phantom with known density to estimate the corresponding spectrum of a CT scanner. To determine the x-ray spectrum of a CT scanner, energy-resolved detectors such as cadmium zinc telluride (CdZnTe)~\citep{miyajima2002,fritz2011}, cadmium telluride (CdTe)~\citep{miyajima2003,redus2009} or high purity germanium~\citep{fewell1981} are often used to directly measure the spectra. However, due to the limited count rate of these detectors and the high photon flux of CT scanners, it is not easy to directly measure the spectrum of a CT scanner because of the detector pile-up effect. Besides, direct measurements using these energy-resolved detectors may suffer from environmental requirement (such as low temperature requirement) or hole trapping effects, which yields low-energy tailing of the energy spectrum~\citep{koenig2012}. Instead of direct measurement, Compton scattering techniques can be employed to reconstruct the spectrum~\citep{yaffe1976,matscheko1987,Bartol2013}. In this method, an energy spectrometer is placed outside the primary ray to receive Compton scattering photons from a scatter object with significantly reduced photon events. The incident photon energy was then calculated with the corresponding energy of the Compton scattering photons and the scattering angle.

Alternatively, the properties of a polychromatic spectrum can be extracted from projection measurements. For example, a harder spectrum will generally yield less attenuated projection measurements and a softer spectrum more attenuated projection measurements. Thus, an alternative solution for spectral measurement is to estimate the spectra with transmission measurements~\citep{sidky2005,zhang2007,duan2011,zhang2013}. These methods usually employ a step or a wedge phantom and a polychromatic forward projection equation formulated as a discrete linear system.  In this linear system, each energy bin of the spectrum is described as an unknown variable. When employing a step wedge phantom, the total number of unknown variables maybe larger than the number of transmission measurements. Thus the linear system may be ill-conditioned and the expectation-maximization (EM) algorithm is often used to solve the problem. In order to further improve the estimation accuracy of the characteristic peak of the spectrum, an improved parameter spectrum model was also employed to recover the spectrum~\citep{zhang2013}.

In this work, we propose a new spectrum estimation method based on indirect transmission measurements and model spectra mixture approach. Instead of using a dedicated step or wedge phantom for transmission measurements, this method aims to calculate the propagation path length (PPL) for each of the segmented materials for each detector pixel and the raw projection data (after taking a logarithm) is compared to the polychromatic reprojection data which is calculated using the PPL and an estimated spectrum. The estimated spectrum is then iteratively updated to minimize the difference of the raw projection data and the polychromatic reprojection data of the volumetric images. In order to make the iterative algorithm robust and stable, the estimated spectrum was expressed as the weighted summation of a set of model spectra which significantly reduce the degrees of freeom (DOF) of the spectrum estimation problem. The model spectra can be generated with either Monte Carlo simulation or other analytical spectrum generators. Spectra recovered with experimental phantom data were compared against the spectrum generated using SpekCalc software~\citep{poludniowski2009} with x-ray source configurations provided by the scanner manufacturer.
%

\section{METHODS}
The problem of spectrum estimation based on transmission measurements is primarily an inverse problem. In order to determine the spectrum, we have to determine the content of each energy bin of the spectrum using the raw projection data. Depending on the width of the energy bin, the inverse problem typically has tens of unknown variables. Namely, there are tens of DOF of the inverse problem. Instead of estimating each energy bin content for a spectrum $\Omega(E)$, a weighted summation of a set of model spectra $\Omega_{i}(E)$ is used to express $\Omega(E)$.
\begin{equation}\label{equ:spek}
\Omega(E)=\sum_{i=1}^{M}c_{i}\Omega_{i}(E).
\end{equation}
Here $M$ is the number of the model spectra and $c_i$ is the weight on the respective model spectrum. The model spectra can be generated with MC simulation tool-kit or any other kinds of spectrum generator, such as SpekCalc~\citep{poludniowski2007a,poludniowski2007b,poludniowski2009}. Now the number of unknown variables of the spectrum estimation problem can be reduced from the number of energy bins to the number of model spectra $M$, i.e. a significant reduction of the DOF of the spectrum estimation problem is achieved by using the model spectra mixture approach.

\begin{figure}[t]
    \includegraphics[width=\textwidth]{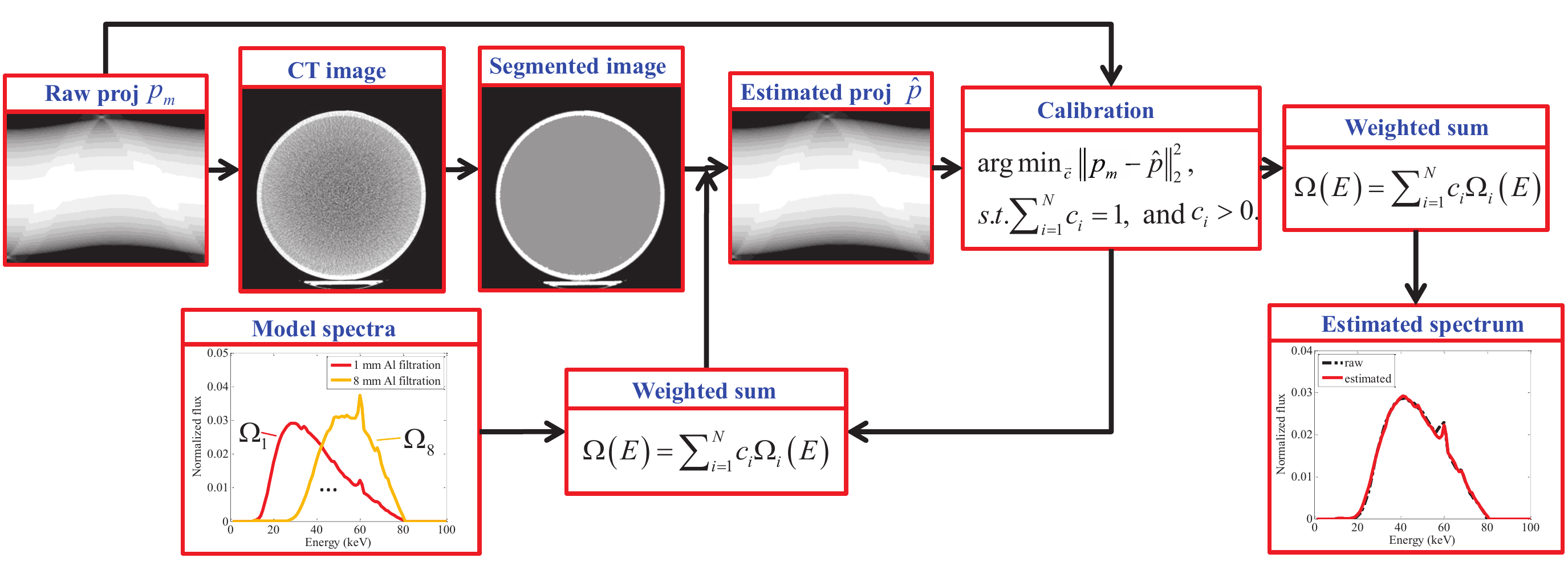}
    \caption{A flowchart of the proposed spectrum estimation method. The method starts with the raw projection data. After an initial CT image reconstruction, the image is segmented into different components, and a polychromatic reprojection is performed with the segmented image and a weighted summation of a set of model spectra. The unknown weights are calibrated by iteratively minimizing the difference of the estimated reprojection data and the raw projection data. The final spectrum is calculated using the weights and the model spectra. }
    \label{fig:f1}
\end{figure}

Figure~\ref{fig:f1} shows the flowchart of the proposed indirect transmission measurements-based spectrum estimation method. The method starts from the original raw projection data. After an initial CT image reconstruction, the image is segmented to different components. The linear attenuation coefficients which can be obtained from the NIST database are assigned to these components. The polychromatic reprojection is then performed using $\Omega(E)$ and the segmented image to generate an estimated projection data. By iteratively updating the weights $c_i$, we can find a set of optimal weights $c_i$ to minimize the difference of the estimated projection data and the raw projection data. The final spectrum is calculated by linearly combining the model spectrum $\Omega_{i}(E)$ with the corresponding weight $c_i$.

\subsection{Polychromatic reprojection}

The polychromatic projection of an object can be represented as
\begin{equation}\label{equ:polyreproj}
\hat{I}=N\int_{0}^{E_{max}}\mathrm{d}E\,\Omega(E) \, \eta(E)\,\mathrm{exp}\left[-\int_{0}^{l}\mu(E,s)\mathrm{d}s\right],
\end{equation}
where $N$ is the total number of photons, $\eta(E)$ is the energy dependent response of the detector and can be considered proportional to photon energy $E$ because most CT scanners use energy-integrating detectors. $E_{max}$ is the maximum photon energy of the spectrum. $\mu(E,s)$ is the energy-dependent linear attenuation coefficient and $l$ is the propagation path length for each ray. Note that $\hat{I}$ is detector pixel dependent and the detector channel index is omitted for convenience. With no object in the beam path, the flood field $I_{0}$  can be expressed as follows:
\begin{equation}\label{equ:floodfield}
\hat{I}_{0}=N\int_{0}^{E_{max}}\mathrm{d}E\,\Omega(E) \, \eta(E).
\end{equation}
After applying the logarithmic operation, the projection data can be expressed as:
\begin{equation}\label{equ:esproj}
\hat{p}=\log\left(\frac{\hat{I}_{0}}{\hat{I}}\right)=\log\left(\frac{\int_{0}^{E_{max}}\mathrm{d}E\,\Omega(E)\,\eta(E)}{\int_{0}^{E_{max}}\mathrm{d}E\,\Omega(E)\,\eta(E)\,\mathrm{exp}\left[-\int_{0}^{l}\mu(E,s)\mathrm{d}s\right]}\right).
\end{equation}
For a given spectrum estimate and segmented image, an estimated polychromatic reprojection dataset $\hat{p}$ can be calculated using equation~(\ref{equ:esproj}). In addition, from the normalization condition, we have
\begin{equation}\label{equ:norm}
\int_{0}^{E_{max}}\mathrm{d}E\,\Omega(E)=1,
\end{equation}
\begin{equation}\label{equ:bNorm}
\int_{0}^{E_{max}}\mathrm{d}E\,\Omega_{i}(E)=1.
\end{equation}
Substituting equation~(\ref{equ:spek}) into equation~(\ref{equ:norm}) and combining with equation~(\ref{equ:bNorm}), we can get the following constraint condition for weights $\mathbf{c}$,
\begin{equation}\label{equ:cons}
\sum_{i=1}^{M}c_{i}=1.
\end{equation}
This constraint will be used to calibrate the weights of the model spectra.

\subsection{Calibrating the weights}

For each detector pixel measurement, $p_m$, the difference between the measurement and the corresponding estimated reprojection, $\hat{p}$, should be minimal if the estimated spectrum matches the actual raw spectrum. We can minimize the difference between $p_m$ and $\hat{p}$ by updating the weights $\mathbf{c}$. Thus the optimal set of $\mathbf{c}$ can be achieved by solving the following optimization problem,
\begin{equation}\label{equ:opt-constraint}
\mathbf{c}= \underset{\mathbf{c}} {\mathrm{argmin}}\;\|p_{m}-\hat{p}\|_{2}^{2}.
\end{equation}

In order to keep the solution of the above equation physically meaningful, a non-negative constraint can also be applied to the weights. Together with the constraint equation~(\ref{equ:cons}), the above unconstrained optimization problem is formulated into a constraint problem as follows,
\begin{equation}\label{equ:opt-constraint2}
\mathbf{c}=\underset{\mathbf{c}} {\mathrm{argmin}}\;\|p_{m}-\hat{p}\|_{2}^{2},    ~~\mathrm{s.t.}~\sum_{i=1}^{M}c_{i}=1,~\mathrm{and}~c_{i}>0.
\end{equation}
To solve the constraint optimization problem of equation~(\ref{equ:opt-constraint2}), in this study, we have reformulated it into an unconstrained problem and solved the unconstrained problem using a simple multi-variable downhill simplex method. Thus the original constraint problem of equation~(\ref{equ:opt-constraint}) can be solved with a sequential optimization approach, minimizing the objective function, followed by normalizing the solution and enforcing non-negative constraint sequentially. The algorithm can be summarized as follows:

\begin{adjustwidth}{0.6cm}{}
\textbf{Algorithm (Sequential optimization)}\\
1. Set k=0, choose $\mathbf{c}_0$.\\
2. \textbf{repeat}\\
3. $\mathbf{c}^{k+1}=\underset{\mathbf{c}} {\mathrm{argmin}}\;\|p_{m}-\hat{p}^k\|_{2}^{2}$\\
4. $\mathbf{c}^{k+1}=\mathbf{c}^{k+1}/\mathrm{sum}(\mathbf{c}^{k+1})$\\
5. $\mathbf{c}^{k+1}=(\mathbf{c}^{k+1})_{+}$\\
6. $k\leftarrow k+1$\\
7. \textbf{until} stopping criterion is satisfied.
\end{adjustwidth}

Line 1 of the algorithm establishes the initial values of the iterative procedure. In this work, the weights for the initial model spectra, $\mathbf{c}_0$, were either set to be equal or to provide the hardest possible spectrum (i.e. the weight of the hardest model spectrum was set to one and all other weights were set to zeros). Lines 2 to 7 iteratively loop to calibrate the optimal weights beginning with the initial weights established in line 1 and terminating when either the maximum iteration number has been met or the difference between the weights of the current and previous iteration ($\mathbf{c}_k$ and $\mathbf{c}_{k+1}$) is smaller than $\varepsilon.$ In this work, the following stopping criterion was used: $\varepsilon=10^{-5}$. Within the loop, line 3 is an inner update of $\mathbf{c}$ to minimize the quadratic error between the measured projection data $p_{m}$ and the estimated reprojection data $\hat{p}$. This is an unconstrained optimization problem and a multi-variable downhill simplex method~\citep{press2007} was employed to solve it. The updated weights, $\mathbf{c},$ are then normalized (line 4) and constrained to be non-negative, with weights $<0$ set equal to 0 (line 5). Finally, the current values are assigned to be the starting point of the next iteration in line 6. The optimal weights generated with the above algorithm are used  to determine the final spectrum.

It should be noted that the original constraint problem equation~(\ref{equ:opt-constraint2}) is not complicated and it can also be solved directly using the methods that can intrinsically handle constraints, such as BFGS-B~\citep{byrd1995} and augmented Lagrangian method (ALM)~\citep{nocedal2006}.

\subsection{Model spectrum generation}

In order to generate the model spectra, the MC simulation tool-kit Geant4~\citep{agostinelli2003} can be employed. Geant4 provides comprehensive physics modelling capabilities embedded in a flexible structure and it offers a complete electron interactions modelling, such as the bremsstrahlung effect and characteristic radiation. With the low energy package G4EMLOW which can describe the electromagnetic interactions of photons, electrons, hadrons and ions with matter down to very low energies (eV scale), Geant4 can precisely model the x-ray photons generation in many kinds of targets and the results of energy spectra have been well validated for keV energy levels~\citep{taschereau2006,miceli2007a,miceli2007,guthoff2012}.

The x-ray source modelled with Geant4 includes an electron gun, a tungsten target, a beryllium window, filters, and a detector. The target angle was set to 20 degrees. Monochromatic electrons were emitted from the electron gun and hit the target to generate photons. The photons propagate through the beryllium window and the filtration material and arrive at the detector. The filtration material used in this work was aluminium. Photon detection was achieved by recording photons (both bremsstrahlung and characteristic X-rays) using the detector located at 300 mm from the anode. The size of the detector was limited to avoid oblique photon events. Aluminium filters of different thicknesses were added to simulate a set of model spectra. 

A sensitive detector was used to store information regarding interactions of a particle, and was attached to the detector volume to extract and store photon energy. A so-called ``sensitive detector'' is a detector model in GEANT4 which tracks particle interactions. Instead of recording the deposited energy of photons within the detector, we directly counted photons and extracted their energy using the sensitive detector. Thus the Geant4-based MC simulation can provide an accurate energy spectrum of the x-ray source by excluding the contribution of the detector, such as K-edge effect of the detector material. For each of the x-ray model spectrum, a total of $5\times10^8$ electrons were simulated. ROOT, an object oriented data analysis framework was used to analysis the recorded photon events. Energy spectrum of x-ray photons for different thicknesses of filtration were recorded as histograms.

In order to estimate the spectrum as accurately as possible, it is recommended that the model spectra should cover the unknown true spectrum. Specifically, the energy range of the model spectra should be at least as broad as that of the true spectrum. For example, if the true spectrum was a 70 kVp polychromatic spectrum filtered with 5 mm of aluminium, then the softest model spectrum should be softer than the true spectrum and the hardest model spectrum should be harder than the true spectrum.

Even though MC simulation can generate unattenuated model spectra which guarantee the actual raw spectrum will be covered, it is not a necessary requirement to generate sufficient model spectra with MC simulations. Other spectrum generators, such as SpekCalc~\citep{poludniowski2007a,poludniowski2007b,poludniowski2009}, can also be used to generate a set of model spectra with a large dynamic range.

\section{Implementation}

The proposed spectrum estimation method was evaluated using both simulated and experimental data. In the following subsections, we will present the implementation details.
\begin{figure}[t]
    \includegraphics[width=\textwidth]{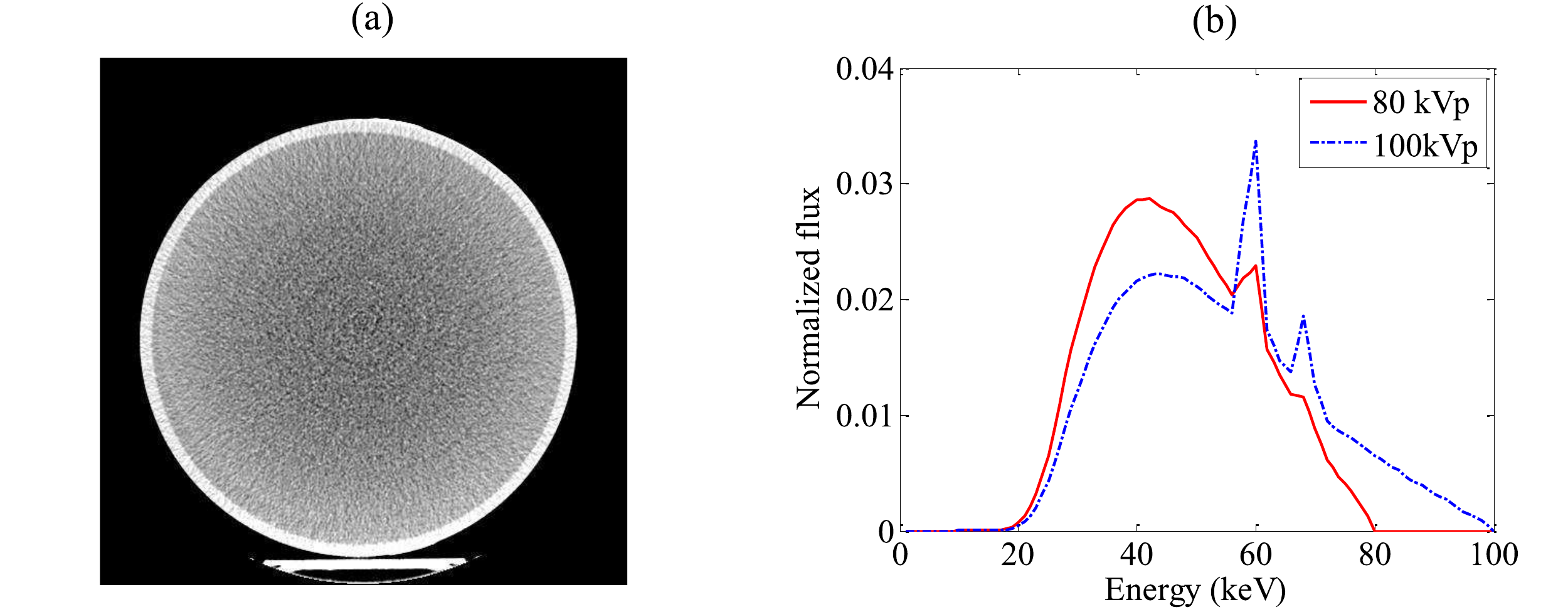}
    \caption{(a) Numerical phantom, (b) the 80 kVp and 100 kVp incident spectra generated using Spektr software.  }
    \label{fig:f2}
\end{figure}

\subsection{Simulation study}
\subsubsection{Simulation settings}

In order to make the numerical simulations as realistic as possible, a numerical phantom was created using a phantom image acquired on a cone beam CT scanner. The acquired water tank image (figure~\ref{fig:f2}) was first segmented to identify the water filling and the PMMA boundary. The patient table was also recognized as PMMA. Raw projection data was generated by polychromatic forward projecting the numerical phantom with an analytical calculated spectrum, purposely differing from the model spectra that will be used in the calibration procedure. Two sets of analytical polychromatic spectra with tube potentials of 100 and 80 kVp were generated using the spectrum generator Spektr~\citep{siewerdsen2004} and both of them were filtered with 3.5 mm aluminium. Since the raw projection data were generated using these two analytical spectra, the two spectra can serve as ground truth for evaluation.

\begin{figure}[t]
    \includegraphics[width=\textwidth]{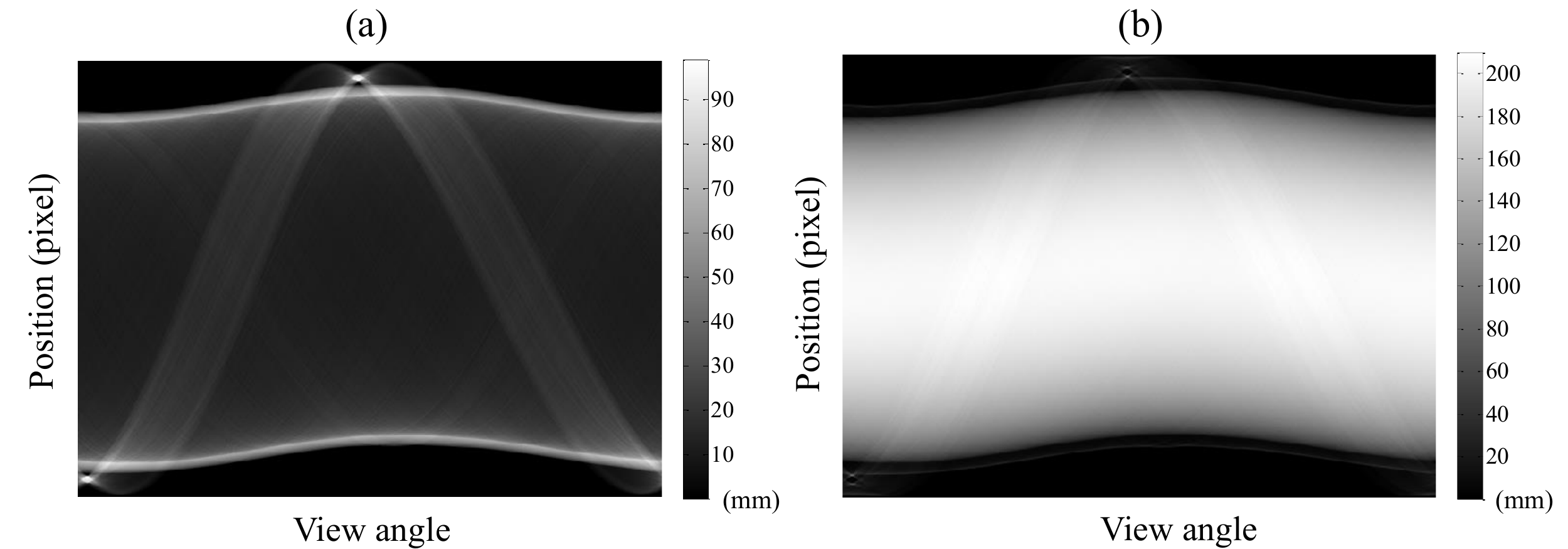}
    \caption{Propagation path length in PMMA (a) and water (b) of the numerical water tank phantom.}
    \label{fig:ppl}
\end{figure}

The source to isocenter distance of the simulated system was 1000 mm and the source to detector distance was 1500 mm. A circular scan which contains 675 projections in 360 degree angular range was simulated. Only one row of detector pixels (the central row) was simulated. The pixel size was 0.776 mm and the detector had 504 pixels. The propagation path length (PPL) for each material for each pixel was calculated using a GPU-based ray tracing algorithm, as illustrated in figure~\ref{fig:ppl}. Projections for each pixel were then calculated with the 100 kVp and 80 kVp analytical spectra and the PPL using the polychromatic projection formula (equation~(\ref{equ:polyreproj})). After taking the logarithmic operation, the projection was regarded as raw projection data. During the spectrum estimation, all of the model spectra are normalized and the softest and the hardest model spectra always cover the raw spectrum for all evaluations. Because we can not yield the correct spectrum if the raw spectrum is out the dynamic range of the model spectra, regardless of how we combine the model spectra.  During the weight calibration procedure, all weights were initialized with the same value.

\subsubsection{Number of model spectra}

In our first evaluation, the number of model spectra that were needed to yield an accurate estimated spectrum was determined and the influence of the number of model spectra on the accuracy of the estimated spectrum was also investigated. For the 100 and 80 kVp raw spectra, spectrum estimations with 2, 4, and 6 model spectra were performed.

In order to quantitatively characterize the accuracy of the estimated spectrum, we use the normalized root mean square error NRMSE and the mean energy difference between the true spectrum and the estimated spectrum $\Delta E$ which were defined as follows:
\begin{equation}\label{equ:error}
NRMSE=\sqrt{\frac{\sum_{e=1}^{N}(\hat{\Omega}(e)-\Omega(e))^{2}}{\sum_{e=1}^{N}\Omega(e)^{2}}}    
\end{equation}
\begin{equation}\label{equ:meanEnergy}
\Delta E= \sum_{e=1}^{N}E_e\,(\Omega(e)-\hat{\Omega}(e))
\end{equation}
Here $\hat{\Omega}(e)$ is the $e$th energy bin of the normalized estimated spectrum and $\Omega(e)$ is $e$th energy bin of the normalized raw spectrum. $N$ is the total number of the energy bins of the spectrum, and $E(e)$ is the energy of the $e$th energy bin.

\subsubsection{Number of phantom materials}

The aim of the proposed method is to estimate the raw spectrum using the corresponding raw projection data. Nevertheless, in realistic applications, it is necessary to handle projections of objects with a varying number of materials. Thus, to make the method as realistic as possible, we also investigated the impact of the number of phantom materials on the accuracy of the estimated spectrum. The numerical phantom was generated using CT images of a water tank. In an initial simulation, both the water filling and the PMMA boundary were considered to be water, thus the numerical phantom only has one material. Then the raw projection data was generated by projecting the numerical phantom with the polychromatic projection formula. In a second simulation, the PMMA boundary was maintained and the raw projection data was generated by a polychromatic projection of the two materials. The raw spectra were estimated with six model spectra in both of the simulations. The accuracy of the estimated spectra was also evaluated with equations~(\ref{equ:error}) and~(\ref{equ:meanEnergy}).

%

\subsubsection{Robustness to model spectra generator}

The model spectra used in the previous simulations were generated using a MC tool-kit. However, compared to other spectrum generators, this process was time consuming and a large number of photon events are needed to yield a spectrum with good statistical properties. Other spectrum calculators, such as SpekCalc~\citep{poludniowski2009}, are much esier to handle and are well validated. Spectrum generation using these generators is very fast (usually on the order of seconds) and this time savings may play an important role in some applications, such as real-time full-spectral image reconstruction~\citep{cai2013}. Therefore, we also investigated the robustness of the proposed method with respect to the model spectra generator.

For the numerical water tank phantom, a 70 kVp analytical raw spectrum which was modelled using Spektr~\citep{siewerdsen2004}, was employed to yield the raw polychromatic projection data. In the spectrum estimation procedure, the model spectra which were generated using both Geant4 and SpekCalc were used to estimate the raw spectrum. The number of model spectra were selected according to the previous simulation results. The accuracy of the estimated spectra were quantitatively characterized using NRMSE and $\Delta E$.

\subsection{Experimental study}

The proposed method was also evaluated using experimental phantom data. The image uniformity module of the Catphan$^{\textregistered}$600\footnote{http://www.phantomlab.com/library/pdf/catphan600\_download.pdf.} phantom which was cast from a uniform material, was used to estimate the corresponding spectra of the acquisition protocols of the CT scanner. The raw projection data was acquired with a clinical bi-plane C-Arm angiography system (Artis zee, Siemens AG, Forchheim, Germany). Table~\ref{tab:paras} shows an overview of the system parameters and acquisition parameters.


\begin{table}[h]
\centering
\caption{System parameters and acquisition parameters of the experimental study.}
\label{tab:paras}
\begin{center}
\begin{tabular}{ll} 
\toprule
\rule[-1ex]{0pt}{3.5ex}  Parameters & Artis zee \\ 
\hline
\rule[-1ex]{0pt}{3.5ex}  Source to detector distance & $1200$ mm \\
\rule[-1ex]{0pt}{3.5ex}  Source to isocenter distance & $749.3$ mm  \\
\rule[-1ex]{0pt}{3.5ex}  Angular range & $198.1$  \\
\rule[-1ex]{0pt}{3.5ex}  View-angle increment & $0.4$  \\
\rule[-1ex]{0pt}{3.5ex}  Number of views & $496$  \\
\rule[-1ex]{0pt}{3.5ex}  X-ray source & $82.5 $ kVp~$\sim90$ kVp   \\
\rule[-1ex]{0pt}{3.5ex}  Cone angle & $8.7^{\circ}$  \\
\rule[-1ex]{0pt}{3.5ex}  Scanning time & 20~s    \\
\rule[-1ex]{0pt}{3.5ex}  Detector size& $380\times296$~mm$^{2}$  \\
\rule[-1ex]{0pt}{3.5ex}  Number of detector pixels & $1232\times960$ (with~$2\times2$~rebinning)  \\
\rule[-1ex]{0pt}{3.5ex}  Detector material& aSi with CsI scintillator \\
\bottomrule
\end{tabular}
\end{center}
\vspace{-1em}
\end{table}

The C-Arm scanners were equipped with auto exposure control (AEC) systems which can perform automatic angular modulation of the tube current, tube potential or both. The modulation was performed according to the object's size, shape and materials distribution to enhance radiation dose efficiency. For the Catphan$^{\textregistered}$600 phantom CT scan, the tube potential was modulated between 82.5 kV and 90 kV. In this case, we can either estimate the spectrum of each view angle by using the corresponding raw projection data of each view angle, or alternatively, estimate an effective spectrum for the whole scan by using raw projections of all view angles. The accuracy of the effective spectrum can be evaluated using full spectrum reconstruction~\citep{cai2013} or artefacts correction to see whether there are residual artefacts. However, this is out the scope of this study. In this study, the spectrum of a single view angle (the corresponding tube potential is 83.8 kV) was estimated independently, and its accuracy was evaluated by comparing the estimated spectrum to the spectrum generated using SpekCalc software~\citep{poludniowski2009} with a beam hardening equivalent of 2.5 mm of aluminium filtration (matching the value provided by the scanner manufacturer).


To reduce the computational complexity with no loss of generality, only the central row of the raw projection data was extracted and compared to the polychromatic reprojection data which was calculated with a projection matrix. The raw spectrum was recovered by updating the weights, $\mathbf{c}$, to minimize the quadratic error between the raw projection data and the reprojection data. During the weights calibration procedure, all of the initial guess were set to the hardest spectrum of the model spectra and the standard NIST attenuation coefficients of water was assigned to the segmented uniform module.

Since the logarithmic raw projection data will be compared to the polychromatic reprojection data which doesn't model the scatter signal, the scatter radiation contribution should be minimized in the raw projection data. In the experimental evaluation, the scanner was equipped with a focused anti-scatter grid and a narrow collimation mode was employed during the CT scan. Therefore, the contribution of the scatter signal to the raw projection was neglected.

Compared to the diagnostic CT scanner, the absorption efficiency of the flat detector equipped on the C-Arm scanner is relatively low, especially for high energy photons. In this evaluation, a spectrum estimation both with and without taking the absorption efficiency into account was performed. After incorporating the absorption efficiency, the energy-dependent response of an energy-integrating detector should be the multiplication of the photon energy and the absorption efficiency. The energy dependent absorption efficiency was calculated using both the analytical Beer's law and MC simulations with different energy levels.

\section{Results}


\subsection{Numerical phantom}

\begin{figure}
    \includegraphics[width=\textwidth]{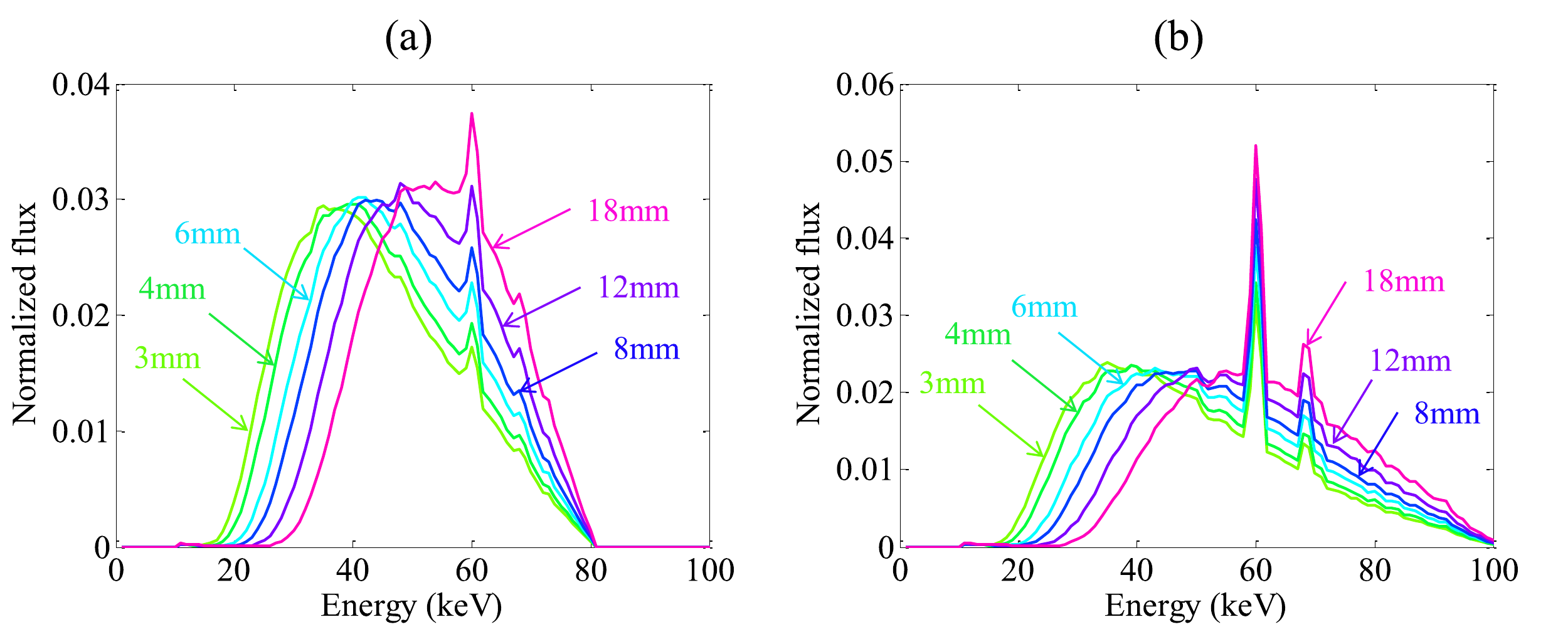}
    \caption{80 (a) and 100 (b) kVp model spectra generated using Geant4 with different thickness of aluminium filtration. }
    \label{fig:f5}
\end{figure}

\subsubsection{Number of model spectra}

\begin{figure}
    \includegraphics[width=\textwidth]{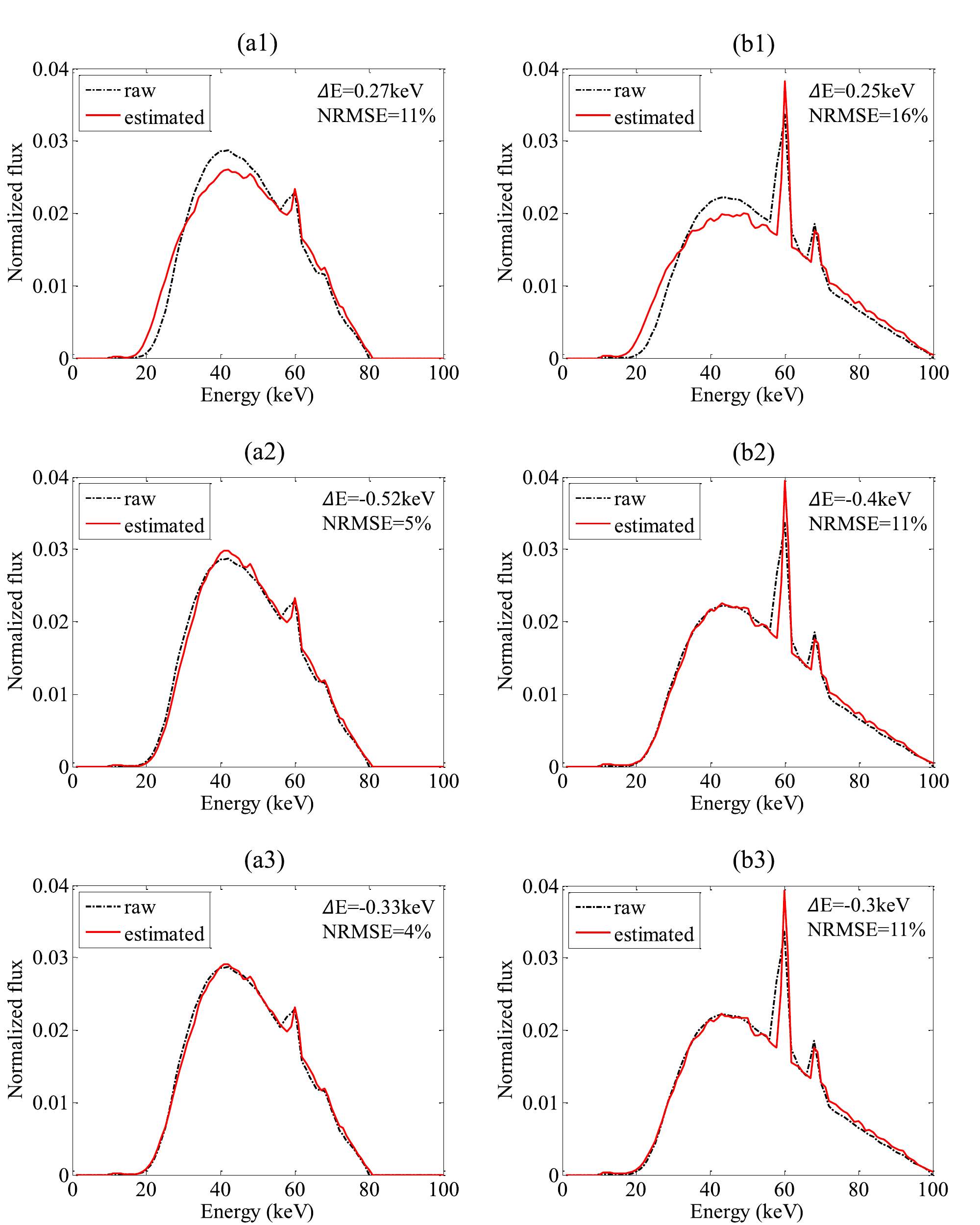}
    \caption{X-ray spectra estimated from the numerical phantom using different number of model spectra. (a1), (a2) and (a3) are the 80 kVp spectra estimated using 2, 4 and 6 Monte Carlo model spectra, respectively. (b1), (b2) and (b3) are the 100 kVp spectra estimated using 2, 4 and 6 Monte Carlo model spectra, respectively. The difference between the estimated spectra and the raw spectra are marginal when six model spectra are used.}
    \label{fig:f6}
\end{figure}

\begin{table}
\vspace{0em}
\caption{Weights of the model spectra of spectrum estimation for the numerical water tank phantom using six model spectra.}
\label{tab:weight}
\begin{center}
\begin{tabular}{ccccccc} 
\toprule
\multicolumn{1}{ c }{\multirow{2}{*}{Tube potential (kV)} } & \multicolumn{6}{ c }{Weights} \\ \cmidrule{2-7}
& $c_1$ & $c_2$ & $c_3$ & $c_4$ & $c_5$ & $c_6$  \\
\hline
\rule[-1ex]{0pt}{3.5ex}  80 & 0.0730 & 0.4023 & 0.2874 & 0.1561 & 0.0493 & 0.0319 \\
\rule[-1ex]{0pt}{3.5ex}  100 & 0.0171 & 0.0369 & 0.9379 & 0.0038 & 0.0023 & 0.0020 \\
\bottomrule
\end{tabular}
\end{center}
\end{table}

Figure~\ref{fig:f5} shows the 80 kVp and 100 kVp model spectra that were used in the numerical evaluation. These model spectra were generated using MC simulation tool-kit Geant4 with different thicknesses of aluminium filters. Figure~\ref{fig:f6} shows the results of the estimated 80 kVp and 100 kVp spectra as a function of the number of MC model spectra, using the numerical water tank phantom. All mean energy differences, $\Delta E$, and normalized root mean square errors, NRMSE, are derived by comparing the estimated spectra with the raw analytical spectra depicted in figure~\ref{fig:f2}(b).

Figure~\ref{fig:f6}(a1), (a2) and (a3) show the 80 kVp spectra estimated with 2, 4 and 6 model spectra and figure~\ref{fig:f6}(b1), (b2) and (b3) show the 100 kVp spectra estimated with 2, 4 and 6 model spectra, respectively. For both of the two cases, NRMSE decreased as the number of model spectra increased, while $\Delta E$ didn't have a significant change with a different number of model spectra. As can be seen, the spectrum estimation with 6 model spectra leads to minor improvements in the accuracy. However, estimation with four model spectra can still yield acceptable results. Table~\ref{tab:weight} shows the calibrated weights of the model spectra of spectrum estimation for the numerical water tank phantom using six model spectra.

NRMSE of the 100 kVp estimated spectra are much larger than NRMSE of the 80 kVp estimated spectra, even though both estimated spectra match the raw analytical spectra quite well. This discrepancy can be attributed to the mismatch of the characteristic peak which is of larger magnitude for the 100 kVp spectrum than the 80 kVp spectrum. The model spectra were generated using Geant4-based MC simulation tool-kit and were binned to 1 keV intervals. However, the raw spectra were generated using the Spektr software which is based on the TASMIP model~\citep{boone1997,sisniega2013}and the TASMIP model is further based completely on interpolation of Fewell's direct spectroscopy measurements results~\citep{fewell1981}, which were tabulated at intervals of 2 keV. Thus, even though the raw Spektr values were interpolated to 1 keV intervals, it is difficult for the estimated spectrum to match the raw analytical spectra at the characteristic peak exactly.

Based on these results, we chose six model spectra such that the NRMSE was minimized for all further simulations. It has to note that the spectra shown in figure~\ref{fig:f5} and~\ref{fig:f6} are estimated using the sequential optimization approach, however, both BFGS-B and ALM methods can provide the similar results. In this study, we use the sequential optimization approach for the sake of simplicity.

\subsubsection{Number of phantom materials}

\begin{figure}
    \includegraphics[width=\textwidth]{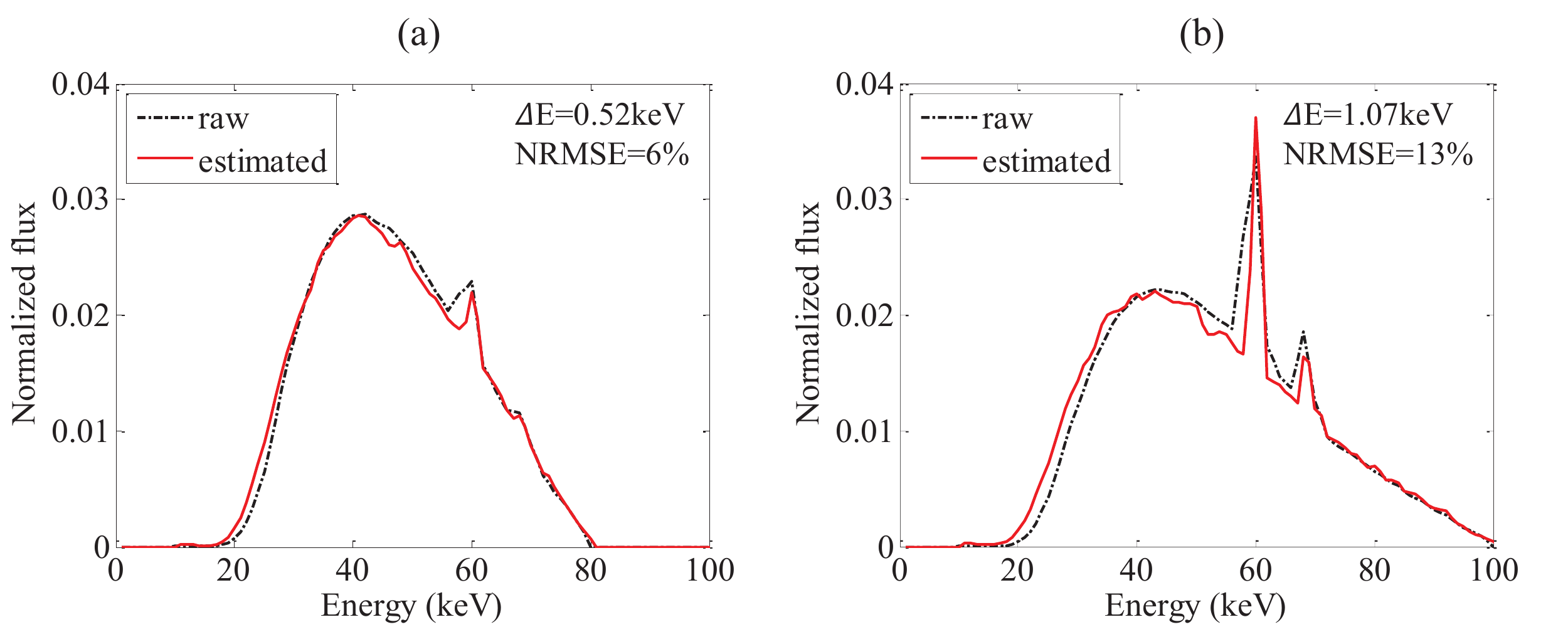}
    \caption{Both 80 (a) and 100 (b) kVp x-ray spectra estimated from the numerical water tank phantom, assuming a single object material by setting the PMMA boundary to water. Compared to the spectra estimated using two materials, the spectra estimated using one material show inferior results.}
    \label{fig:f7}
\end{figure}

In figure~\ref{fig:f7}, the influence of the number of phantom materials used in estimation procedure on accuracy of the estimated spectrum with the proposed method is shown. Figure~\ref{fig:f7}(a) and (b) depict the estimated 80 kVp and 100 kVp spectra with only one material. For better demonstration, both estimations use six MC model spectra. Compared to figure~\ref{fig:f6}(a3) and (b3) which also use six model spectra to estimate the 80 kVp and 100 kVp spectra, the NRMSE increase from $4\%$ to $6\%$ and from $11\%$ to $13\%$ for the 80 kVp and the 100 kVp spectra, respectively. The reason for this behaviour can be attributed to the attenuation property of the polychromatic spectrum within different kinds of materials. If a single material phantom is used in the spectrum estimation, the estimated spectrum only takes the attenuation property of one material into account during the spectrum recovery. However, when a phantom consisting of more than one material is used, the estimated spectrum has to take all of the attenuation properties of these materials into account. This makes the spectrum recovery problem more deterministic. In practice, multi-materials phantom or a multi-materials CT image is recommended for the spectrum estimation problem.


\subsubsection{Robustness to model spectra generator}

\begin{figure}
    \includegraphics[width=\textwidth]{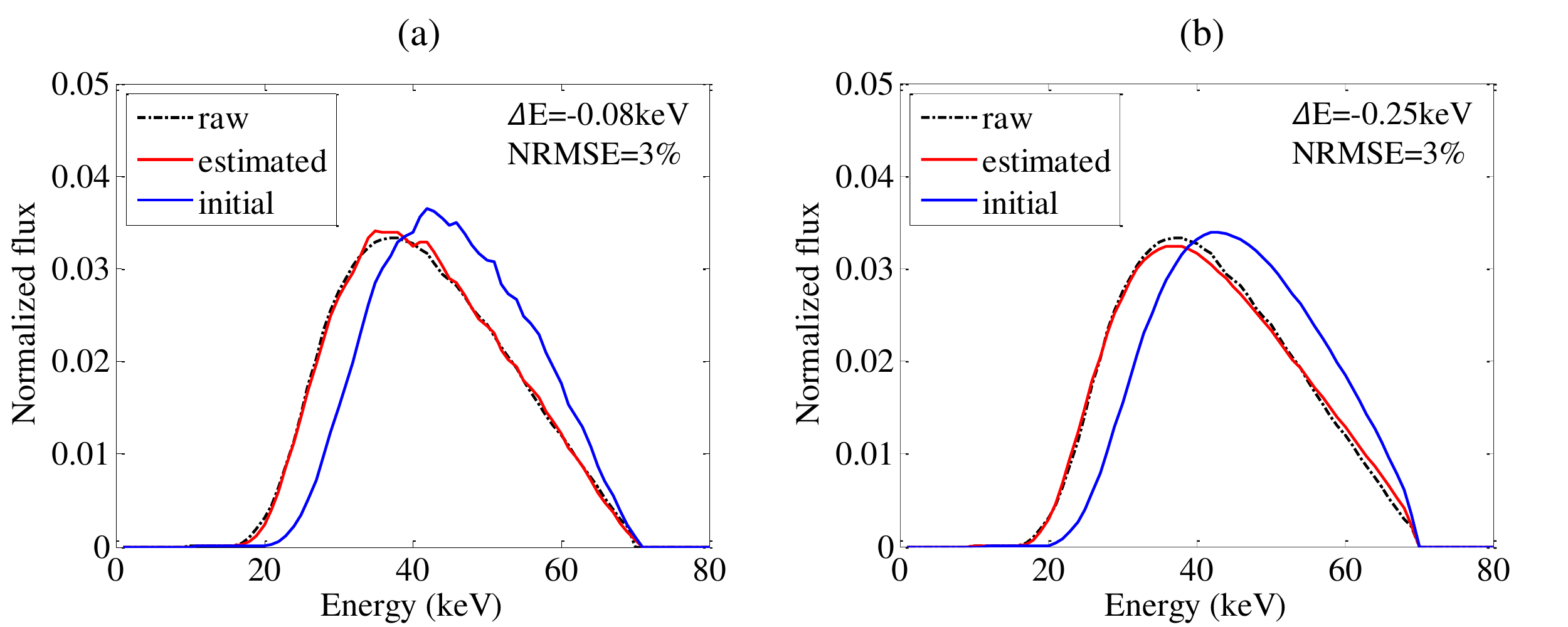}
    \caption{X-ray spectra estimated for the numerical water tank phantom data. Model spectra generated using both Geant4 (a) and SpekCalc (b) were employed in the estimation procedure to demonstrate the proposed method was robust against the model spectra generator. The initial guesses for the spectrum recovery problem correspond to the hardest model spectrum. Spectra estimated using both model spectra yield acceptable results, showing the proposed method was robust with respect to the model spectra generator. }
    \label{fig:rSpekGenerator}
\end{figure}

In order to evaluate the robustness of the proposed method with respect to model spectra generator, both Geant4 model spectra and SpekCalc model spectra were used to recover the raw Spektr analytical spectrum. In figure~\ref{fig:rSpekGenerator}(a), a set of 70 kVp model spectra which were generated using MC tool-kit Geant4 with different thickness of filtration were used in the weights calibration procedure. In figure~\ref{fig:rSpekGenerator}(b), the model spectra were generated using SpekCalc. The initial guess of the iterative calibration procedure was the hardest spectrum of the model spectra. Quantitative analysis of NRMSE and $\Delta E$ show spectra estimated using both model spectra generators can yield acceptable results, indicating the proposed method is robust with respect to the model spectra generator. Note that all of the spectra have 1 keV energy bin widths and are normalized.

\subsection{Experimental phantom data}

\begin{figure}[t]
    \centering
    \includegraphics[width=0.5\textwidth]{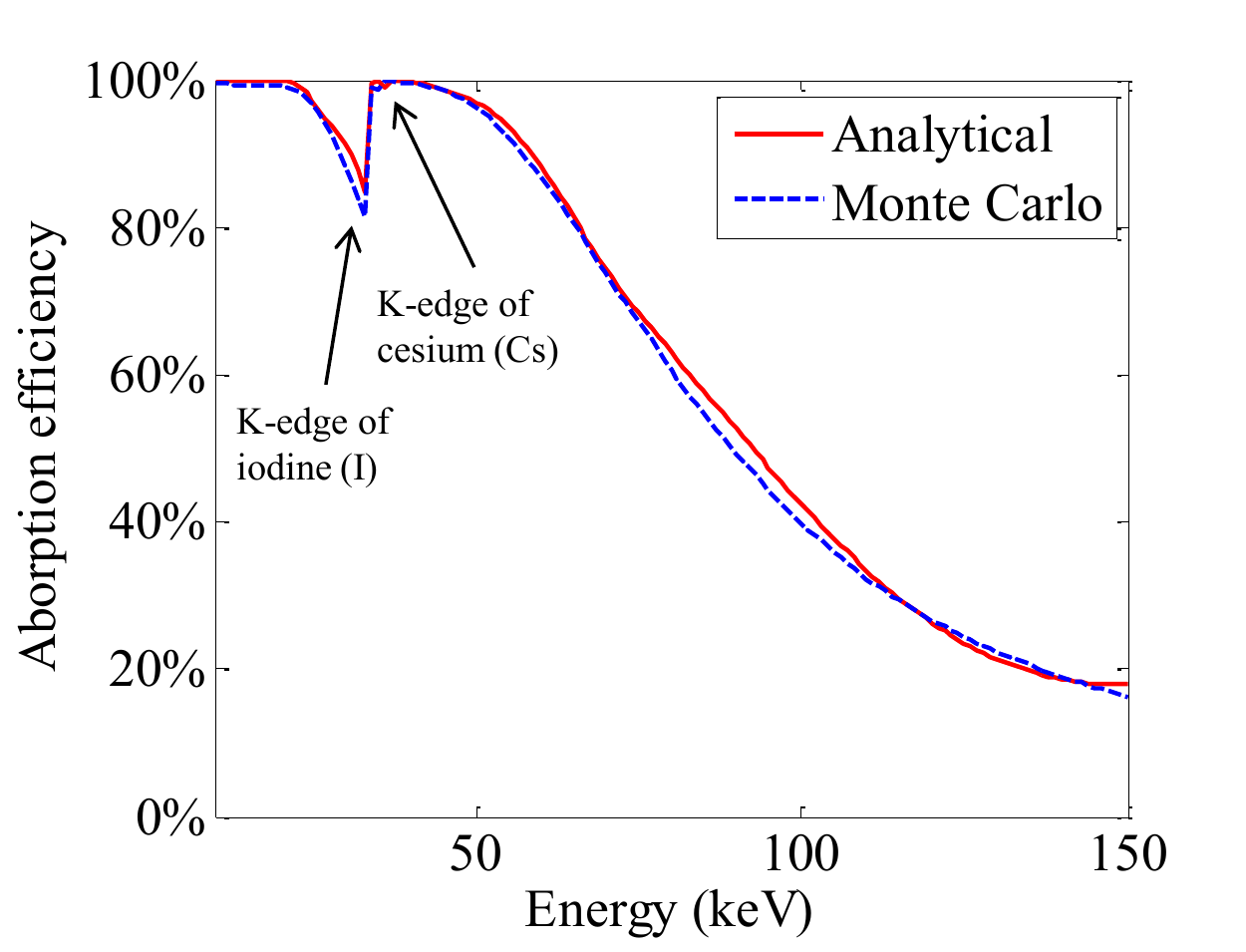}
    \caption{Energy dependent absorption efficiency of the CsI scintillator calculated using both the analytical method and MC simulation. Usually, the absorption efficiency goes down as photon energy increases, the absorption efficiency around 100\% for low energies, however the K-edges of Cs and I elevate the absorption efficiency at 36.0 keV and 33.2 keV, respectively. }
    \label{fig:absorptionEff}
\end{figure}

Spectrum estimation using the proposed method with physical phantom was also performed on a clinical bi-plane C-Arm system (Artis zee, Siemens AG, Forchheim, Germany). Because of the low absorption efficiency of the flat detector that was equipped on the C-Arm scanner, we need to incorporate the absorption efficiency in the polychromatic reprojection procedure. Figure~\ref{fig:absorptionEff} shows the energy dependent absorption efficiency of the flat detector with 0.6 mm thickness of CsI. Both analytical calculation and MC simulation results are depicted. Usually, the absorption efficiency goes down as photon energy increases. However, the K-edges of Cs and I elevates the absorption efficiency at 36.0 keV and 33.2 keV, respectively.

\begin{figure}[t]
    \centering
    \includegraphics[width=\textwidth]{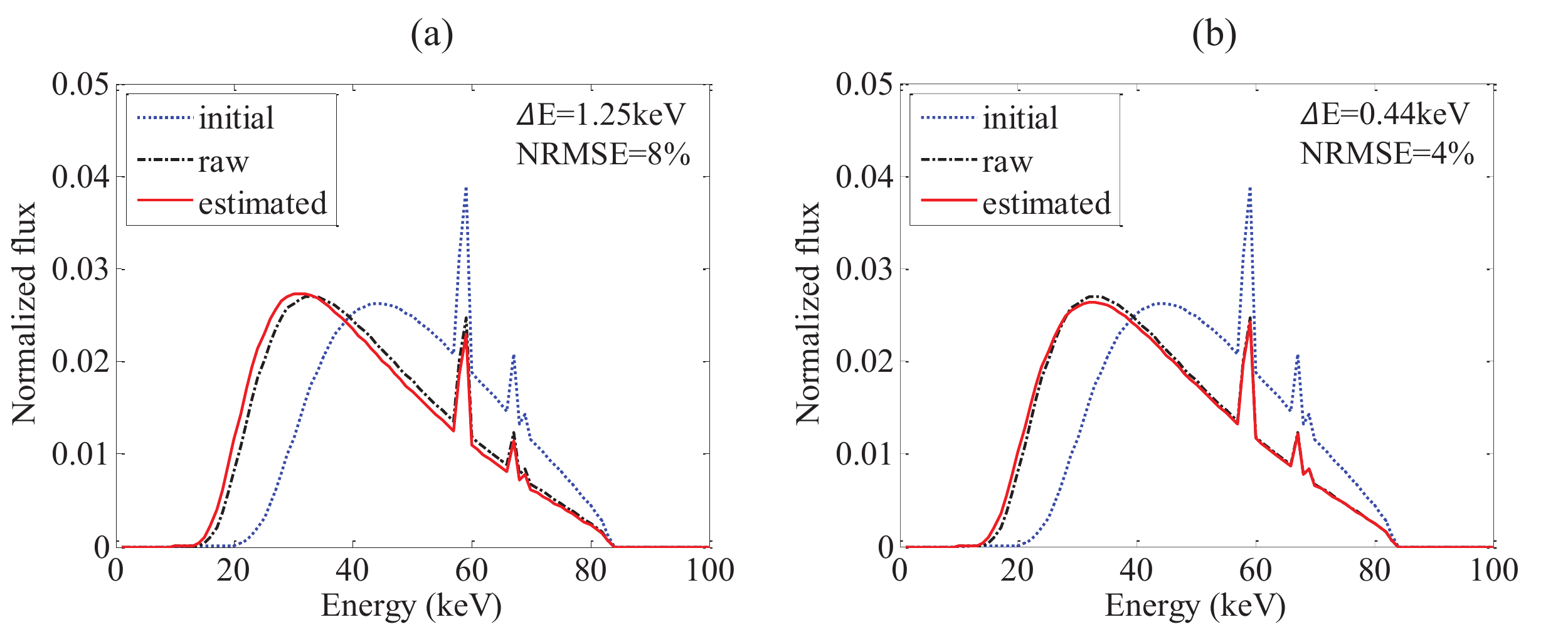}
    \caption{X-ray spectra estimated for the C-Arm system using the Catphan$^{\textregistered}$600 phantom data with (a) and without (b) absorption efficiency incorporation. Model spectra generated using SpekCalc were employed in the estimation procedure, and the raw spectrum was generated using SpekCalc software and x-ray tube specifications provided by the scanner manufacturer. The initial guesses of the spectrum recovery problem correspond to the hardest model spectrum. Due to the low absorption efficiency, the estimated spectrum can not match the raw spectrum well without taking the absorption efficiency into account for flat panel detector based cone-beam CT, but by incorporating the absorption efficiency, the estimated spectrum can be greatly improved.}
    \label{fig:catphan}
\end{figure}

By taking the energy dependent absorption efficiency into account, the detector response is described as the multiplication of the photon energy and the absorption efficiency for an energy-integrating detector. Figure~\ref{fig:catphan} shows the estimated 83.8 kVp spectrum by using the Catphan$^{\textregistered}$600 phantom data with and without the incorporation of absorption efficiency. For this experimental phantom study, all of the model spectra used in the weights calibration, were generated by the SpekCalc software. Figure~\ref{fig:catphan}(a) shows the spectrum estimated without taking the absorption efficiency into account. The reference raw spectrum was generated using SpekCalc software with x-ray tube specifications provided by the scanner manufacturer. The initial guess was the hardest model spectrum. As can be seen, the mean energy of the estimated spectrum was 1.25 keV lower than the raw spectrum. This was because the experimental raw projection data tended to yield a lower photon count due to the low absorption efficiency. However, the absorption efficiency was not taken into account (i.e., it was regarded as 100\%) in the polychromatic reprojection procedure. Thus, to compensate the difference of the raw projection data and the polychromatic reprojection data, the estimated spectrum tended to be softer. By taking the photon absorption efficiency into account (figure~\ref{fig:catphan}(b)), the estimated spectrum matched the raw spectrum much better. Both of NRMSE and $\Delta E$ were greatly reduced.

\section{Discussion}
\label{sec:discussion}
The proposed method was based on transmission measurement, namely, by updating the estimated spectrum to minimize the difference between the raw projection data and the polychromatic reprojection data. In this study, the estimated spectrum was expressed as a weighted summation of a set of model spectra. Instead of directly estimating the content of each energy bin of the spectrum, the proposed spectrum estimation method estimates the corresponding weights for each model spectrum. Thus the number of unknown variables for the spectrum recovery problem can be greatly reduced from the number of energy bins to the number of model spectra and the solution of inverse problem was very stable, especially when two materials phantom were employed in the weights calibration procedure. Quantitative evaluations show the proposed method yields acceptable spectra with four model spectra, and the accuracy may be further improved with an increased number of model spectra. Since the proposed method has shown robustness with respect to model spectra generator and it is very convenient to generate spectra with different filtrations using SpekCalc, it is recommended to use a higher number of model spectra in the spectrum recovery procedure, such as six or eight model spectra.

The rationale of a higher number of model spectra yield improved estimation results, is that the model spectra make up a database and the spectrum calibration procedure is to find an optimal spectrum using the database to generate a reprojection dataset whose difference is as small as possible from the raw projection data. Thus the accuracy of estimated spectrum could improve as the sufficiency of the database increases. When only two model spectra(the softest and the hardest model spectra, i.e., a small database) were used, it is difficult for these two spectra to make up a medium filtered spectrum if the raw spectrum is a medium filtered spectrum, as shown in figure~\ref{fig:f6}(a1) and (b1). In this case, the estimated spectrum is inferior due to the insufficiency of the model spectra database. When six model spectra were used, the estimated spectra were significantly improved since it is much easier to find a spectrum using the database which has a similar medium filtered profile (shown in figure~\ref{fig:f6}(a3) and (b3)). One may want to use a large amount of model spectra (a large database) in the weights calibration, however, the calibration time may increase as the number of model spectra increases.

In order to keep the model spectra covering the incident spectrum, one may use a broad filtration during the model spectra generation. It has to note that since the spectrum is the weighted summation of a set of model spectra, the model spectra do impact on the final result. For example, spectrum estimation using Spektr~\citep{siewerdsen2004} model spectra and SpekCalc~\citep{poludniowski2009} model spectra may have different profiles, even though both of the results may have similar mean energies and have minimal NRMSEs. Because most of the widely used spectrum generators are well verified, spectrum estimation using these generators would yield acceptable results.

Since the proposed method tries to minimize the quadratic error of the raw projection data and the estimated reprojection data. The contribution of the raw projection data includes not only the attenuation of the object, but also the extra attenuation (such as bow-tie filter, detector sensor protection materials). Thus the estimated spectrum which was used to calculate the estimated reprojection data, should be an effective spectrum which has taken these extra attenuation into account. In addition, the method described here is localized; in other words, the estimated spectrum corresponds to those detector pixels which provide its raw projection data for the spectrum estimate. Since the central detector row of the raw projection data was used to calibrate the spectrum, the estimated spectrum can be considered to be an effective spectrum for the whole fan beam. The heel effect of the x-ray tube is not taken into account here. However, to estimate an average spectrum along different cone angle which incorporates the average attenuation of the heel effect, raw projection data of the central column of the detector may be employed. Although the spectra along different fan angle and cone angle are different, an average spectrum or effective spectrum may be sufficient for most the realistic spectrum applications.

Because of its pixel-wise feature, the proposed method can be applied to estimate the spectrum of an CT scanner equipped with auto exposure control (AEC) system. To enhance radiation dose efficiency, CT scan with AEC system perform automatic angular modulation of the tube current, tube potential or both from view by view by according to the object's size, shape and materials distribution. Thus the x-ray spectrum may change from view by view during the CT scan. To estimate the spectrum of a specific view angle, the raw projection data of this specific view angle can be employed to estimate the corresponding spectrum. In addition, by using the model spectra of different tube potential and projections from all of the view angles, the proposed method can be further applied to estimate an effective spectrum for a whole CT scan with tube potential modulation. In this case, the correctness of the estimated effective spectrum has to be evaluated using either a full-spectrum image reconstruction, artefact correction or weighted spectra for all view angles. Our aim here was not to investigate the correctness of the estimated effective spectrum but to evaluate the proposed spectrum estimation method. Therefore, the evaluation of the correctness of the effective spectrum of all view angles is not within the scope of this paper. In future studies, we will investigate how to estimate an effective spectrum and to evaluate its correctness using either artefacts correction or weighted spectra from all view angles.

It is important to exclude scattered radiation in the raw projection data. The proposed process compares the raw projection data with the polychromatic reprojection data which does not contain scatter signal. The experimental data based on C-Arm measurements arguably contains a comparably high scatter fraction. However, this did not show considerable influence on the results. This might be attribute to the focused scatter grid employed by the systems, rejecting a considerable amount of scattered radiation. Further work is necessary to test the proposed method to the influence of scattered radiation.


Compared to the diagnostic CT scanner, the energy dependent absorption efficiency of the flat-panel detectors (FPD) that were used in the C-Arm systems is relatively low because the stopping power of CsI-based FPD pixel is inferior than the $\mathrm{Gd_2O_2S}$-based diagnostic CT detector pixel and the thickness of the CsI scintillator is also limited. This is the reason that there was clear discrepancy between the spectrum estimated with and without photon absorption efficiency incorporation for the experimental evaluation. Since the absorption efficiency is much higher for the diagnostic CT scanner, to our belief, even without taking the absorption efficiency into account, spectrum estimation for the diagnostic CT scanner may yield acceptable result.

\section{Conclusions}

We proposed an indirect transmission measurement-based spectrum estimation method that uses the raw projection data of a phantom with known density to recover the corresponding effective spectrum. The method has been evaluated using both simulated data and experimental data. Quantitative analysis demonstrated the method can estimate the x-ray spectra quite well. Further tests showed that a higher number of model spectra and multi-component phantoms or objects can yield improved spectrum estimation. In addition, the proposed method is robust with respect to the model spectra generator, thus several widely used spectra generators (such as Spektr and SpekCalc) can be employed to estimate the spectrum.

\ack
This work is partially supported by a grant from Siemens Medical Solutions. The authors would like to thank Dr. Guanghong Chen for suggesting this work and Dr. Stephen Brunner for providing the water tank images. The authors thank John Garrett for his careful proof-reading of the manuscript and his useful comments. The authors are also grateful to Drs. Ke Li, Yinghua Tao, Tim Szczykutowicz and Yinsheng Li, Yongshuai Ge for helpful discussions.


\section*{References}
\begin{harvard}


\bibitem[Agostinelli {\em et~al.}(2003)Agostinelli, Allison, Amako,
  Apostolakis, Araujo, Arce, Asai, Axen, Banerjee, Barrand, {\em
  et~al.}]{agostinelli2003}
Agostinelli, S., Allison, J., Amako, K.~e., Apostolakis, J., Araujo, H., Arce,
  P., Asai, M., Axen, D., Banerjee, S., Barrand, G., {\em et~al.} (2003).
\newblock Geant4—a simulation toolkit.
\newblock {\em Nuclear instruments and methods in physics research section A:
  Accelerators, Spectrometers, Detectors and Associated Equipment\/}, {\bf
  506}(3), 250--303.

\bibitem[Archer and Wagner(1982)Archer and Wagner]{archer1982}
Archer, B.~R. and Wagner, L.~K. (1982).
\newblock A laplace transform pair model for spectral reconstruction.
\newblock {\em Medical physics\/}, {\bf 9}(6), 844--847.

\bibitem[Ay {\em et~al.}(2004)Ay, Shahriari, Sarkar, Adib, and Zaidi]{ay2004}
Ay, M., Shahriari, M., Sarkar, S., Adib, M., and Zaidi, H. (2004).
\newblock Monte carlo simulation of x-ray spectra in diagnostic radiology and
  mammography using mcnp4c.
\newblock {\em Physics in medicine and biology\/}, {\bf 49}(21), 4897.

\bibitem[Bazalova and Verhaegen(2007)Bazalova and Verhaegen]{bazalova2007}
Bazalova, M. and Verhaegen, F. (2007).
\newblock Monte carlo simulation of a computed tomography x-ray tube.
\newblock {\em Physics in medicine and biology\/}, {\bf 52}(19), 5945.

\bibitem[Bartol(2013)Bartol]{Bartol2013}
Bartol, L.~J. (2013).
\newblock {\em Spectroscopic characterization of high-energy and high fluence
  rate photon beams\/}.
\newblock Ph.D. thesis, University of Wisconsin-Madison.

\bibitem[Birch and Marshall(1979)Birch and Marshall]{birch1979}
Birch, R. and Marshall, M. (1979).
\newblock Computation of bremsstrahlung x-ray spectra and comparison with
  spectra measured with a ge (li) detector.
\newblock {\em Physics in Medicine and Biology\/}, {\bf 24}(3), 505.

\bibitem[Boone and Seibert(1997)Boone and Seibert]{boone1997}
Boone, J.~M. and Seibert, J.~A. (1997).
\newblock An accurate method for computer-generating tungsten anode x-ray
  spectra from 30 to 140 kv.
\newblock {\em Medical physics\/}, {\bf 24}(11), 1661--1670.

\bibitem[Byrd {\em et~al.}(1995)Byrd, Lu, Nocedal, and Zhu]{byrd1995}
Byrd, R.~H., Lu, P., Nocedal, J., and Zhu, C. (1995).
\newblock A limited memory algorithm for bound constrained optimization.
\newblock {\em SIAM Journal on Scientific Computing\/}, {\bf 16}(5),
  1190--1208.

\bibitem[Cai {\em et~al.}(2013)Cai, Rodet, Legoupil, and
  Mohammad-Djafari]{cai2013}
Cai, C., Rodet, T., Legoupil, S., and Mohammad-Djafari, A. (2013).
\newblock A full-spectral bayesian reconstruction approach based on the
  material decomposition model applied in dual-energy computed tomography.
\newblock {\em Medical physics\/}, {\bf 40}(11), 111916.

\bibitem[Duan {\em et~al.}(2011)Duan, Wang, Yu, Leng, and McCollough]{duan2011}
Duan, X., Wang, J., Yu, L., Leng, S., and McCollough, C.~H. (2011).
\newblock Ct scanner x-ray spectrum estimation from transmission measurements.
\newblock {\em Medical physics\/}, {\bf 38}(2), 993--997.

\bibitem[Fewell {\em et~al.}(1981)Fewell, Shuping, and Hawkins]{fewell1981}
Fewell, T.~R., Shuping, R.~E., and Hawkins, K.~R. (1981).
\newblock {\em Handbook of computed tomography x-ray spectra\/}.
\newblock US Department of Health and Human Services, Public Health Service,
  Food and Drug Administration, Bureau of Radiological Health.

\bibitem[Fritz {\em et~al.}(2011)Fritz, Shikhaliev, and Matthews~II]{fritz2011}
Fritz, S.~G., Shikhaliev, P.~M., and Matthews~II, K.~L. (2011).
\newblock Improved x-ray spectroscopy with room temperature czt detectors.
\newblock {\em Physics in medicine and biology\/}, {\bf 56}(17), 5735.

\bibitem[Gallardo {\em et~al.}(2004)Gallardo, R{\'o}denas, and
  Verd{\'u}]{gallardo2004}
Gallardo, S., R{\'o}denas, J., and Verd{\'u}, G. (2004).
\newblock Monte carlo simulation of the compton scattering technique applied to
  characterize diagnostic x-ray spectra.
\newblock {\em Medical physics\/}, {\bf 31}(7), 2082--2090.

\bibitem[Guthoff {\em et~al.}(2012)Guthoff, Brovchenko, de~Boer, Dierlamm,
  M{\"u}ller, Ritter, Schmanau, and Simonis]{guthoff2012}
Guthoff, M., Brovchenko, O., de~Boer, W., Dierlamm, A., M{\"u}ller, T., Ritter,
  A., Schmanau, M., and Simonis, H.-J. (2012).
\newblock Geant4 simulation of a filtered x-ray source for radiation damage
  studies.
\newblock {\em Nuclear Instruments and Methods in Physics Research Section A:
  Accelerators, Spectrometers, Detectors and Associated Equipment\/}, {\bf
  675}, 118--122.

\bibitem[Hassler {\em et~al.}(1998)Hassler, Garnero, and Rizo]{hassler1998}
Hassler, U., Garnero, L., and Rizo, P. (1998).
\newblock X-ray dual-energy calibration based on estimated spectral properties
  of the experimental system.
\newblock {\em Nuclear Science, IEEE Transactions on\/}, {\bf 45}(3),
  1699--1712.

\bibitem[Koenig {\em et~al.}(2012)Koenig, Schulze, Zuber, Rink, Butzer, Hamann,
  Cecilia, Zwerger, Fauler, Fiederle, {\em et~al.}]{koenig2012}
Koenig, T., Schulze, J., Zuber, M., Rink, K., Butzer, J., Hamann, E., Cecilia,
  A., Zwerger, A., Fauler, A., Fiederle, M., {\em et~al.} (2012).
\newblock Imaging properties of small-pixel spectroscopic x-ray detectors based
  on cadmium telluride sensors.
\newblock {\em Physics in medicine and biology\/}, {\bf 57}(21), 6743.

\bibitem[Lin {\em et~al.}(2014)Lin, Ramirez-Giraldo, Gauthier, Stierstorfer,
  and Samei]{lin2014}
Lin, Y., Ramirez-Giraldo, J.~C., Gauthier, D.~J., Stierstorfer, K., and Samei,
  E. (2014).
\newblock An angle-dependent estimation of ct x-ray spectrum from rotational
  transmission measurements.
\newblock {\em Medical physics\/}, {\bf 41}(6), 062104.

\bibitem[Llovet {\em et~al.}(2003)Llovet, Sorbier, Campos, Acosta, and
  Salvat]{llovet2003}
Llovet, X., Sorbier, L., Campos, C., Acosta, E., and Salvat, F. (2003).
\newblock Monte carlo simulation of x-ray spectra generated by
  kilo-electron-volt electrons.
\newblock {\em Journal of applied physics\/}, {\bf 93}(7), 3844--3851.

\bibitem[Maeda {\em et~al.}(2005)Maeda, Matsumoto, and Taniguchi]{maeda2005}
Maeda, K., Matsumoto, M., and Taniguchi, A. (2005).
\newblock Compton-scattering measurement of diagnostic x-ray spectrum using
  high-resolution schottky cdte detector.
\newblock {\em Medical physics\/}, {\bf 32}(6), 1542--1547.

\bibitem[Mainegra-Hing and Kawrakow(2006)Mainegra-Hing and
  Kawrakow]{mainegra2006}
Mainegra-Hing, E. and Kawrakow, I. (2006).
\newblock Efficient x-ray tube simulations.
\newblock {\em Medical physics\/}, {\bf 33}(8), 2683--2690.

\bibitem[Matscheko and Carlsson(1989)Matscheko and
  Carlsson]{matscheko1989compton}
Matscheko, G. and Carlsson, G.~A. (1989).
\newblock Compton spectroscopy in the diagnostic x-ray energy range: I.
  spectrometer design.
\newblock {\em Physics in medicine and biology\/}, {\bf 34}(2), 185.

\bibitem[Matscheko and Ribberfors(1987)Matscheko and Ribberfors]{matscheko1987}
Matscheko, G. and Ribberfors, R. (1987).
\newblock A compton scattering spectrometer for determining x-ray photon energy
  spectra.
\newblock {\em Physics in medicine and biology\/}, {\bf 32}(5), 577.

\bibitem[Matscheko {\em et~al.}(1989)Matscheko, Carlsson, and
  Ribberfors]{matscheko1989}
Matscheko, G., Carlsson, G.~A., and Ribberfors, R. (1989).
\newblock Compton spectroscopy in the diagnostic x-ray energy range: Ii.
  effects of scattering material and energy resolution.
\newblock {\em Physics in medicine and biology\/}, {\bf 34}(2), 199.

\bibitem[Miceli {\em et~al.}(2007a)Miceli, Thierry, Bettuzzi, Flisch, Hofmann,
  Sennhauser, and Casali]{miceli2007a}
Miceli, A., Thierry, R., Bettuzzi, M., Flisch, A., Hofmann, J., Sennhauser, U.,
  and Casali, F. (2007a).
\newblock Comparison of simulated and measured spectra of an industrial 450kv
  x-ray tube.
\newblock {\em Nuclear Instruments and Methods in Physics Research Section A:
  Accelerators, Spectrometers, Detectors and Associated Equipment\/}, {\bf
  580}(1), 123--126.

\bibitem[Miceli {\em et~al.}(2007b)Miceli, Thierry, Flisch, Sennhauser, Casali,
  and Simon]{miceli2007}
Miceli, A., Thierry, R., Flisch, A., Sennhauser, U., Casali, F., and Simon, M.
  (2007b).
\newblock Monte carlo simulations of a high-resolution x-ray ct system for
  industrial applications.
\newblock {\em Nuclear Instruments and Methods in Physics Research Section A:
  Accelerators, Spectrometers, Detectors and Associated Equipment\/}, {\bf
  583}(2), 313--323.

\bibitem[Miyajima(2003)Miyajima]{miyajima2003}
Miyajima, S. (2003).
\newblock Thin cdte detector in diagnostic x-ray spectroscopy.
\newblock {\em Medical physics\/}, {\bf 30}(5), 771--777.

\bibitem[Miyajima {\em et~al.}(2002)Miyajima, Imagawa, and
  Matsumoto]{miyajima2002}
Miyajima, S., Imagawa, K., and Matsumoto, M. (2002).
\newblock Cdznte detector in diagnostic x-ray spectroscopy.
\newblock {\em Medical physics\/}, {\bf 29}(7), 1421--1429.

\bibitem[Nocedal and Wright(2006)Nocedal and Wright]{nocedal2006}
Nocedal, J. and Wright, S.~J. (2006).
\newblock Numerical optimization 2nd, chaper 17.

\bibitem[Poludniowski {\em et~al.}(2009)Poludniowski, Landry, DeBlois, Evans,
  and Verhaegen]{poludniowski2009}
Poludniowski, G., Landry, G., DeBlois, F., Evans, P., and Verhaegen, F. (2009).
\newblock Spekcalc: a program to calculate photon spectra from tungsten anode
  x-ray tubes.
\newblock {\em Physics in medicine and biology\/}, {\bf 54}(19), N433.

\bibitem[Poludniowski(2007)Poludniowski]{poludniowski2007b}
Poludniowski, G.~G. (2007).
\newblock Calculation of x-ray spectra emerging from an x-ray tube. part ii.
  x-ray production and filtration in x-ray targets.
\newblock {\em Medical physics\/}, {\bf 34}(6), 2175--2186.

\bibitem[Poludniowski and Evans(2007)Poludniowski and Evans]{poludniowski2007a}
Poludniowski, G.~G. and Evans, P.~M. (2007).
\newblock Calculation of x-ray spectra emerging from an x-ray tube. part i.
  electron penetration characteristics in x-ray targets.
\newblock {\em Medical physics\/}, {\bf 34}(6), 2164--2174.

\bibitem[Press(2007)Press]{press2007}
Press, W.~H. (2007).
\newblock {\em Numerical recipes 3rd edition: The art of scientific
  computing\/}.
\newblock Cambridge university press.

\bibitem[Redus {\em et~al.}(2009)Redus, Pantazis, Pantazis, Huber, and
  Cross]{redus2009}
Redus, R.~H., Pantazis, J.~A., Pantazis, T.~J., Huber, A.~C., and Cross, B.~J.
  (2009).
\newblock Characterization of cdte detectors for quantitative x-ray
  spectroscopy.
\newblock {\em Nuclear Science, IEEE Transactions on\/}, {\bf 56}(4),
  2524--2532.

\bibitem[Sidky {\em et~al.}(2005)Sidky, Yu, Pan, Zou, and Vannier]{sidky2005}
Sidky, E.~Y., Yu, L., Pan, X., Zou, Y., and Vannier, M. (2005).
\newblock A robust method of x-ray source spectrum estimation from transmission
  measurements: Demonstrated on computer simulated, scatter-free transmission
  data.
\newblock {\em Journal of Applied Physics\/}, {\bf 97}(12), 124701.

\bibitem[Siewerdsen {\em et~al.}(2004)Siewerdsen, Waese, Moseley, Richard, and
  Jaffray]{siewerdsen2004}
Siewerdsen, J., Waese, A., Moseley, D., Richard, S., and Jaffray, D. (2004).
\newblock Spektr: A computational tool for x-ray spectral analysis and imaging
  system optimization.
\newblock {\em Medical physics\/}, {\bf 31}(11), 3057--3067.

\bibitem[Sisniega {\em et~al.}(2013)Sisniega, Desco, and Vaquero]{sisniega2013}
Sisniega, A., Desco, M., and Vaquero, J. (2013).
\newblock Modification of the tasmip x-ray spectral model for the simulation of
  microfocus x-ray sources.
\newblock {\em Medical physics\/}, {\bf 41}(1), 011902.

\bibitem[Taschereau {\em et~al.}(2006)Taschereau, Chow, Cho, and
  Chatziioannou]{taschereau2006}
Taschereau, R., Chow, P., Cho, J., and Chatziioannou, A. (2006).
\newblock A microct x-ray head model for spectra generation with monte carlo
  simulations.
\newblock {\em Nuclear Instruments and Methods in Physics Research Section A:
  Accelerators, Spectrometers, Detectors and Associated Equipment\/}, {\bf
  569}(2), 373--377.

\bibitem[Tucker {\em et~al.}(1991)Tucker, Barnes, and Chakraborty]{tucker1991}
Tucker, D.~M., Barnes, G.~T., and Chakraborty, D.~P. (1991).
\newblock Semiempirical model for generating tungsten target x-ray spectra.
\newblock {\em Medical physics\/}, {\bf 18}(2), 211--218.

\bibitem[Waggener {\em et~al.}(1999)Waggener, Blough, Terry, Chen, Lee, Zhang,
  and McDavid]{waggener1999}
Waggener, R.~G., Blough, M.~M., Terry, J.~A., Chen, D., Lee, N.~E., Zhang, S.,
  and McDavid, W.~D. (1999).
\newblock X-ray spectra estimation using attenuation measurements from 25 kvp
  to 18 mv.
\newblock {\em Medical physics\/}, {\bf 26}(7), 1269--1278.

\bibitem[Xu {\em et~al.}(2009)Xu, Langan, Wu, Pack, Benson, Tkaczky, and
  Schmitz]{xu2009}
Xu, D., Langan, D.~A., Wu, X., Pack, J.~D., Benson, T.~M., Tkaczky, J.~E., and
  Schmitz, A.~M. (2009).
\newblock Dual energy ct via fast kvp switching spectrum estimation.
\newblock In {\em SPIE Medical Imaging\/}, pages 72583T--72583T. International
  Society for Optics and Photonics.

\bibitem[Yaffe {\em et~al.}(1976)Yaffe, Taylor, and Johns]{yaffe1976}
Yaffe, M., Taylor, K., and Johns, H. (1976).
\newblock Spectroscopy of diagnostic x rays by a compton-scatter method.
\newblock {\em Medical physics\/}, {\bf 3}(5), 328--334.

\bibitem[Zhang and Zhang(2013)Zhang and Zhang]{zhang2013}
Zhang, H. and Zhang, P. (2013).
\newblock Model-based x-ray spectrum estimation from scanning data of ct
  phantoms.
\newblock In {\em The 12th International Meeting on Fully Three-Dimensional
  Image Reconstruction in Radiology and Nuclear Medicine, Granlibakken Resort,
  Lake Tahoe, CA, June 16-21\/}, pages pp. 181--184.

\bibitem[Zhang {\em et~al.}(2007)Zhang, Zhang, Chen, Xing, Cheng, and
  Xiao]{zhang2007}
Zhang, L., Zhang, G., Chen, Z., Xing, Y., Cheng, J., and Xiao, Y. (2007).
\newblock X-ray spectrum estimation from transmission measurements using the
  expectation maximization method.
\newblock In {\em Nuclear Science Symposium Conference Record, 2007. NSS'07.
  IEEE\/}, volume~4, pages 3089--3093. IEEE.



\end{harvard}

\end{document}